\documentclass[reprint,prb,amsmath,amssymb,groupedaddress]{revtex4-1}

\usepackage{graphicx}
\usepackage{dcolumn}
\usepackage{bm}
\usepackage{color}

\begin{document}

\title{Magnetic nanotubes: A new material platform to realize robust spin-Seebeck effect and perfect thermal spin-filtering effect}

\author{Dan-Dan Wu$^{1}$}
\author{Hua-Hua Fu$^{1,2,}$}
\email{hhfu@mail.hust.edu.cn}
\author{Qing-Bo Liu$^{1}$}
\author{Gui-Fang Du$^{1}$}
\author{Ruqian Wu$^{2},$}%
 \altaffiliation{wur@uci.edu}
\affiliation{%
$^1$School of Physics and Wuhan National High Magnetic field center,
Huazhong University of Science and Technology, Wuhan 430074,
People's Republic of China.\\
$^2$Department of Physics and Astronomy, University of California, Irvine, California 92697-4575, United States.}%

\date{\today}

\begin{abstract}
To construct reliable material platforms and to uncover new rules to realize spin-Seebeck effect (SSE) and thermal spin-filtering effect (SFE) are core topics in spin caloritronics. Here we design several single-layer boron-nitrogen nanotubes (BNNTs) with \emph{n} boron (nitrogen) atoms substituted by carbons in every unit cell. We find that for \emph{n} = 1, the magnetic BNNTs generate a good SSE with nearly symmetric spin-up and spin-down currents; while as the carbon dopant concentration increases (c.f. \emph{n} $\geq$ 2), a high rotational symmetry of the carbons contributes to generate
the SSE with more symmetric thermal spin-up and spin-down currents, otherwise towards the thermal SFE. Moreover, some metallic BNNTs can generate the SSE or the SFE with finite threshold temperatures, due to the compensation effect around the Fermi level. More importantly, we find that the compression strain engineering is an effective route to improve these effects and to realize the transition between them. These theoretical results about the SSE in nanotubes enrich the spin caloritronics, and put forwards new material candidates to realize the SSE and other inspiring thermospin phenomena.

\begin{description}
\item[PACS number(s)]
73.63.-b, 73.23.Hk, 75.47.-m, 75.75.-c
\end{description}

\end{abstract}

\pacs{72.15.Rn, 72.25.-b, 73.20.-r, 78.67.De}

\maketitle

\section{INTRODUCTION}

The spin-Seebeck effect (SSE) \cite{1,2,3,4,5}, referred to as the generation of spin current by a temperature gradient in ferromagnetic junctions, has been a central topic of spin caloritronics research, because it provides a new opportunity for design of energy-efficient nanodevices. It has already established that spin-Seebeck currents can be carried by either spin waves or conduction electrons \cite{6}. The former is the most important mechanism in insulating magnets where spin angular momenta are carried by spin wave or magnon propagation \cite{7,8,9}. In the latter case, a temperature gradient is applied in magnetic metals or semiconductors to generate opposite flow direction of thermal electrons in the two spin channels \cite{10,11,12}, which was first confirmed by observing a thermal spin current at the Cu/NiFe interface \cite{13}. For the development of recent nanotechnologies, ferromagnetic nanoribbons (FNs) have been used as model systems for the realization of the SSE and other thermal spin effects \cite{14,15,16,17,18}.

In FNs, the thermal spin-up and spin-down currents usually transport along two opposite directions along the edges \cite{19,20,21}. This is not conductive to the cancellation of charge currents to reduce the Joule heat. A pure spin-Seebeck current, i.e., the thermal charge current is zero, is achievable if the transport channels of spin-up and spin-down electrons overlap fully in space. To this end, one needs to reduce the nanoribbon widths, but this usually scarifies the structural stabilities. Furthermore, the spatial separation of the spin-up and spin-down currents makes the SSE difficult to manipulate. These disadvantages suggest the need to explore more efficient and stable candidates for the realization of the SSE based on conduction electrons.

It is well known that nanotubes are exceptional materials in which many excellent physical properties are integrated together \cite{22}. Recent research works have demonstrated that finite-layer nanotubes are promising thermoelectric materials with many advantages \cite{22,23,24}. First, nanotubes are one-dimensional materials, leading to high Seebeck coefficient due to the quantum confinement effect \cite{25,26,27}. Second, the natural network structures with multiple intertube junctions significantly reduce the thermal conductivity, leading to large figure-of-merit values \cite{28,29}. Finally, in comparison with the nanoribbon structures, nanotubes are usually mechanically more strong and flexible, which ensures the structural stability in thermal spin device applications, even in attachment with curved and movable objects \cite{30,31}. To realize the SSE and obtain pure thermal spin currents in nanotubes, however, is still a task. One of the reasons is that pristine nanotubes are mostly nonmagnetic. It is encouraging that nanotubes can become magnetic by including defects and dopants, or via hydrogenation \cite{32,33,34}.

\begin{figure*}
\centering
\includegraphics[width=5.5in]{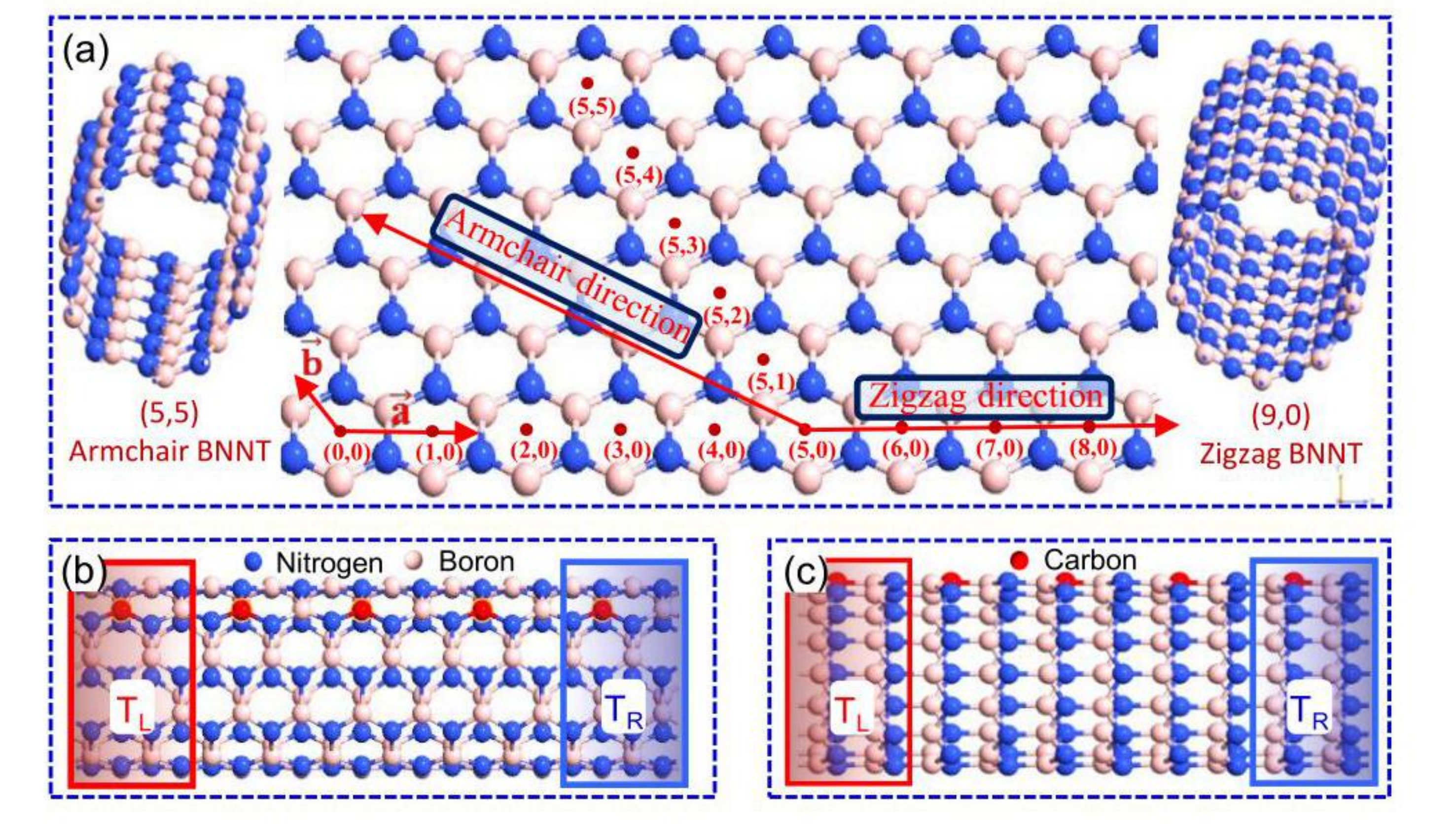}
\caption{\label{fig:wide} (a) A monatomic layer of boron-nitrogen (BN) and two kinds of BN nanotubes. Vectors $\vec{a}$ and $\vec{b}$ are unit vectors of the two-dimensional lattice. The coordinates of the lattice point are marked in the figure, and are useful to distinguish the crystal structures of tubules. The left and the right panels plot the pure (5,5) armchair boron-nitrogen nanotube (BNNT) and (9,0) zigzag BNNT. (b) and (c) The schematic structures of spin caloritronic devices based on the one-carbon-substituted (5,5) armchair BNNT and (9,0) zigzag BNNT in their unit cells. The temperatures in the source and the drain are $T_L$ and $T_R$, respectively, and the temperature gradient between the source and the drain can be defined as $\Delta$$T$ ($= T_L-T_R$).}
\end{figure*}

In this work, we construct several single-layer boron-nitrogen nanotubes (BNNTs), and substitute some boron (or nitrogen) atoms in one boron-nitrogen (BN) ring of the unit cell by carbon to induce magnetization. By performing theoretical calculations, we find some practical rules to achieve the SSE and other thermal spin effects in magnetic nanotube systems. If the number of boron atoms to be substituted in the BN ring is set as \emph{n}, we find that for $n = 1$, both armchair and zigzag carbon-substituted BNNTs can generate a good SSE with nearly symmetric spin-up and spin-down currents, while for $n \geq 2$, the rotational symmetry of the substituted carbon atoms in the BN ring determines the thermal spin transport properties: a high rotational symmetry of the carbons about the BN ring contributes to generate an enhanced SSE with more symmetric spin-up and spin-down currents, otherwise tends to thermal spin-filtering effect (SFE) with the spin polarization high to 100$\%$. Moreover, some metallic and magnetic carbon-substituted BNNTs can also generate the SSE or thermal SFE with finite threshold temperature ($T_{th}$), due to the compensation effect of particles and holes around the Fermi level. Subsequently, we apply the compression strain engineering to improve and control the SSE and the SFE, to realize the transition between them, and even to obtain other inspiring thermo-spin transport phenomena, such as the temperature-dependent SFE. Finally, we put forwards an effective route to make sure that the Curie temperature ($T_c$) should be larger than $T_{th}$ of the magnetic nanotubes, which is regarded as a required condition to ensure the occurrence of the thermal spin effects.

The remainder of this paper is organized as follows. In Sec. II, we construct two types of magnetic carbon-substituted nanotubes based on single-layer BN sheets, design the thermo-spin devices and introduce the theoretical method in detail. In Sec. III, the numerical results of electronic structures, spin-resolve transmission and thermally driven spin currents of different magnetic BNNTs are presented, some practical rules for the magnetic nanotubes to generate the SSE and the thermal SFE are discussed, and then an effective route to improve the SSE and to realize the transition between the SSE and the SFE is proposed. Finally, a summary is given in the last section.

\section{DEVICE MODELS AND THEORETICAL METHOD}

Considering that general nanotubes with small diameters \emph{d} can exhibit high thermopower and Seebeck coefficient due to the quantum size effect\cite{26,35}, here we use the (5,5)-BNNT (armchair) with $d$ = 0.706 nm and the (9,0)-BNNT (zigzag) with $d$ = 0.718 nm, shown in Fig. 1(a) as model systems. We first confirm that the pristine BNNTs are nonmagnetic insulators. To produce magnetization, we substitute boron atoms in the every unit cells with carbon atoms, as the examples illustrated in Figs. 1(b) and 1(c). Note that the boron (nitrogen) substituted BNNTs have already been fabricated successfully in experiments. For example, Wei \emph{et al.} prepared the carbon-substituted BNNTs by using electron-beam-induced doping strategy and confirmed that this class of nanotubes can exist stably even at room temperature\cite{36}. To illustrate these two spin caloritronic effects, we should focus on how to generate spin currents by a temperature gradient $\Delta$\emph{T} in the thermal spin devices based on the magnetic BNNTs (see Figs. 1(b) and 1(c)). In both devices, the left and right contacts are constructed by the same nanotube, and hold the source temperature $T_{L}$ and the drain one $T_R$, respectively.

Our calculations are performed by using the density-functional theory (DFT) and the non-equilibrium Green's function approach (NEGF) \cite{37,38}, as implemented in the SIESTA code \cite{39} and the TRANSAMPA code \cite{40,41}. A vacuum space larger than 20 {\AA} is placed around the nanotubes in order to avoid the interaction between periodic images. The generalized gradient approximation (GGA) in the form of Perdew-Burke-Ernzerhof (PBE) exchange-correlation function is adopted \cite{42}. The core electrons are represented by norm-conserving pseudo-potentials and the valence electrons are described by a numerical atomic-orbital basis set. The cutoff energy is set as 75 Ry for all tasks. The maximum force tolerance on each atom with 0.02 eV/{\AA} and a $2\times2\times6$ Monkhorst-Pack \emph{k}-point grids in the Brillouin zone are adopted to perform structural optimization. In the Landauer-B\"{u}ttiker formalism, the current is given by the following equation \cite{43}
\begin{align}
I_\sigma=\frac{e}{h}\int_{\infty}^{-\infty}\{T^{\sigma}(E)[f_{L}(E,\mu_{L},T_{L})-f_{R}(E,\mu_{R},T_{R})]\}dE,
\end{align}
where $e$ is the electronic charge, $h$ is the Planck¡¯s constant and $\sigma$ = $\uparrow$ or $\downarrow$ indicates the spin index. $f_{\alpha}=\{1+\textrm{exp}[(E-\mu_{\alpha})/{{k_B}}T_{\alpha}]\}^{-1}$ is the Fermi distribution function of contact $\alpha$ with $\alpha$ = $L, R$, where $T_{\alpha}$ and $\mu_{\alpha}$ are the temperature and the chemical potential of contact $\alpha$, respectively. The spin-resolved transmission coefficients of the nanotubes can be calculated by using the NEGF in the linear-resonance regime as\cite{44}
\begin{align}
T^{\sigma}(E)=Tr[\Gamma_{L}G^{R}(E)\Gamma_{R}G^{A}(E)]^{\sigma},
\end{align}
where $\Gamma_{L/R}=i[\Sigma_{L/R}-\Sigma^{\dag}_{L/R}]$ indicates the interaction between a central scattering area and the left/right contact, whose self-energy is $\Sigma_{L/R}$. $G^{R/A}(E)$ represents the retarded (advanced) Green¡¯s function of the central region, $G^{R}(E)=[H_{cen}-(E+i\eta)+\Sigma_{L}+\Sigma_{R}]$ and $G^{A}(E)=[G^{R}(E)]^{\dag}$, where $H_{cen}$ is the Hamiltonian in the central scattering region. Moreover, the spin-resolved transmission coefficient $T^{\sigma}(E)$ can also be calculated through the elements of the scattering matrix of the devices \cite{43,44}. Considering that the magnetic nanotubes we adopted are structurally perfect, electrons can move freely in the systems. Since there is no scattering in the nanotube homojunctions, the spin-resolved transmission coefficient is reduced to the number of spin transmission channels across the contacts. In the numerical calculations of the spin transport through the devices, the number of spin-dependent transport channels can be determined by counting the number of Bloch waves which propagate along a given direction \cite{45}.Here, we focus on the spin-dependent currents driven by the temperature gradient in the nanotubes, without the presence of any external bias and back voltages. Because of $|\Delta{T}| \ll T_{L(R)}$, the system produces a small potential difference $\Delta{\mu} \ll T_{L(R)}$.

\section{RESULTS AND DISCUSSION}

Firstly, the supercells of the armchair and zigzag BNNTs are constituted by 40 and 36 atoms respectively, in which one boron atom is substituted by carbon as shown in Figs. 1(b) and 1(c). The DFT calculations show that both nanotubes are spin semiconducting, and the energy bands are symmetric about the Fermi level ($E_F = 0$). It is noted that the symmetric spin-splitting bands maintain if the size of the supercells is changed, as shown in supplemental Fig. S1 \cite{46}. The specific bands lead to the symmetric spin-dependent transmission spectra plotted in Figs. 2(c) and 2(d). When a temperature gradient $\Delta$\emph{T} is applied between the source and the drain, $f_L (T_L) - f_R (T_L-\Delta{T})$ is longer equal to zero and shows inverse symmetric behaviors around the Fermi level\cite{45}. Since $f_L - f_R$ is an odd function, the sign of $I_\sigma$ is determined by the slop of the transmission coefficient $T_\sigma$ near the Fermi level according to Eq. (1)\cite{16,47}. Considering that the spin-up and spin-down transmission spectra are located below and above the Fermi level respectively, we expect that spin-up current ($I_{up}$) and the spin-down one ($I_{dn}$) have opposite signs. Indeed, one may see that $I_{dn} > 0$ and $I_{up} < 0$ from the DFT calculation, as plotted in Figs. 2(e) and 2(f). Interesting, the amplitudes of $I_{up}$ and $I_{dn}$ are the same in the entire temperature range. This indicates that the charge current $I_c$ $(=I_{up} + I_{dn})$ is suppressed to zero and hence a nearly pure thermal spin current $I_{s}$ $(=\frac{h}{2e}(I_{up} - I_{dn}))$ is produced. Thus a satisfactory SSE, characterized by the same threshold temperature ($T_{th}$) in $I_{up}$ and $I_{dn}$, is achieved here, which can be considered as the first practical rule for the carbon-substituted BNNT to generate the SSE.

\begin{figure}
\includegraphics[width=3.35in]{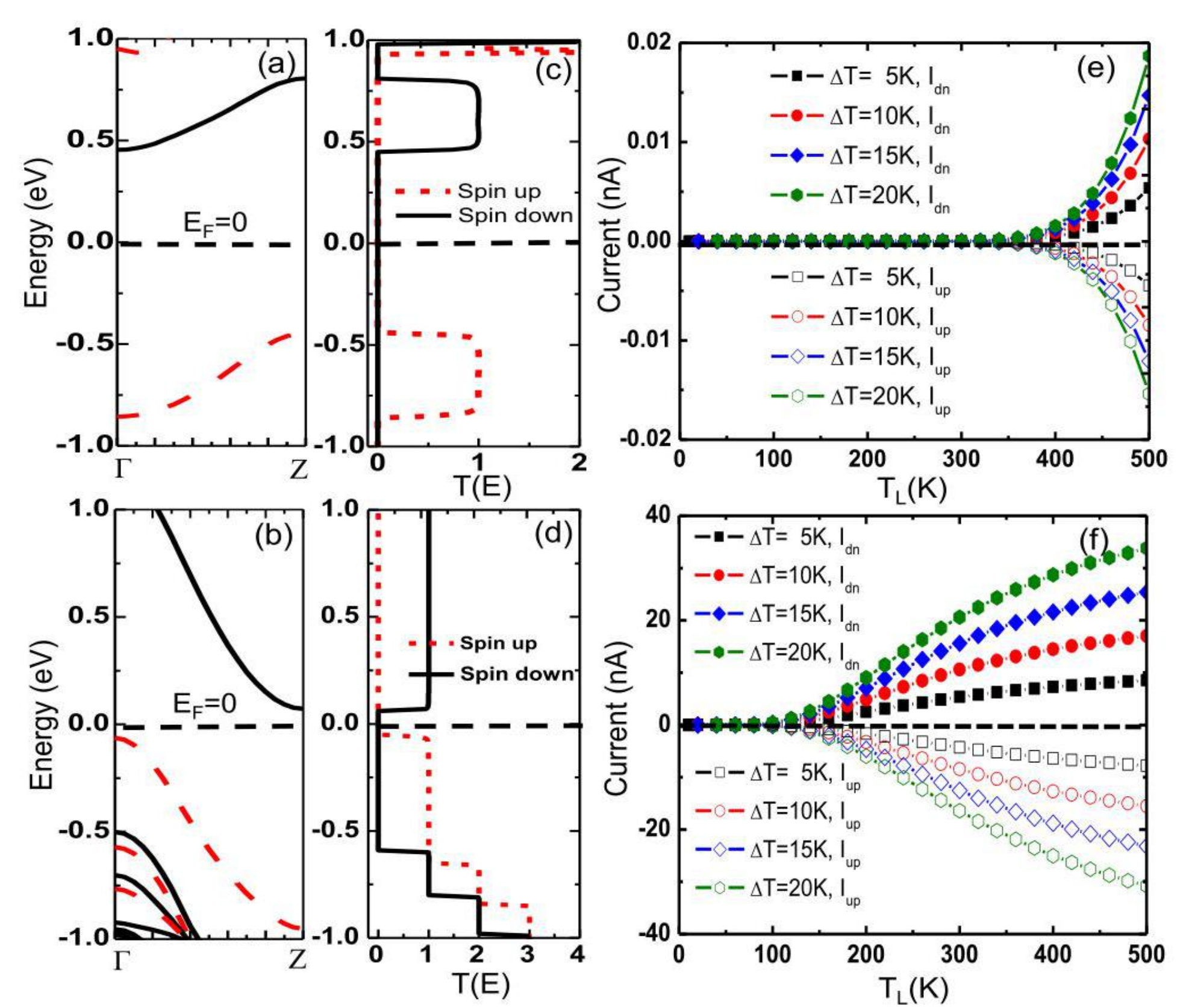}
\caption{\label{fig:wide} (a) and (b) The spin-dependent band structures of the one-carbon-substituted (5,5) armchair and (9,0) zigzag BNNTs, the every unit cells in these two nanotubes are constituted by 40 and 36 atoms (including one carbon atom), respectively. (c) and (d) The corresponding spin-dependent transmissions of these two nanotubes. (e) and (f) The thermally driven spin-up current ($I_{up}$) and spin-down one ($I_{dn}$) as a function of the device temperature $T_L$, where the temperature gradient $\Delta$\emph{T} is set as 5, 10, 15 and 20 K.}
\end{figure}

\begin{figure*}
\includegraphics[width=5.35in]{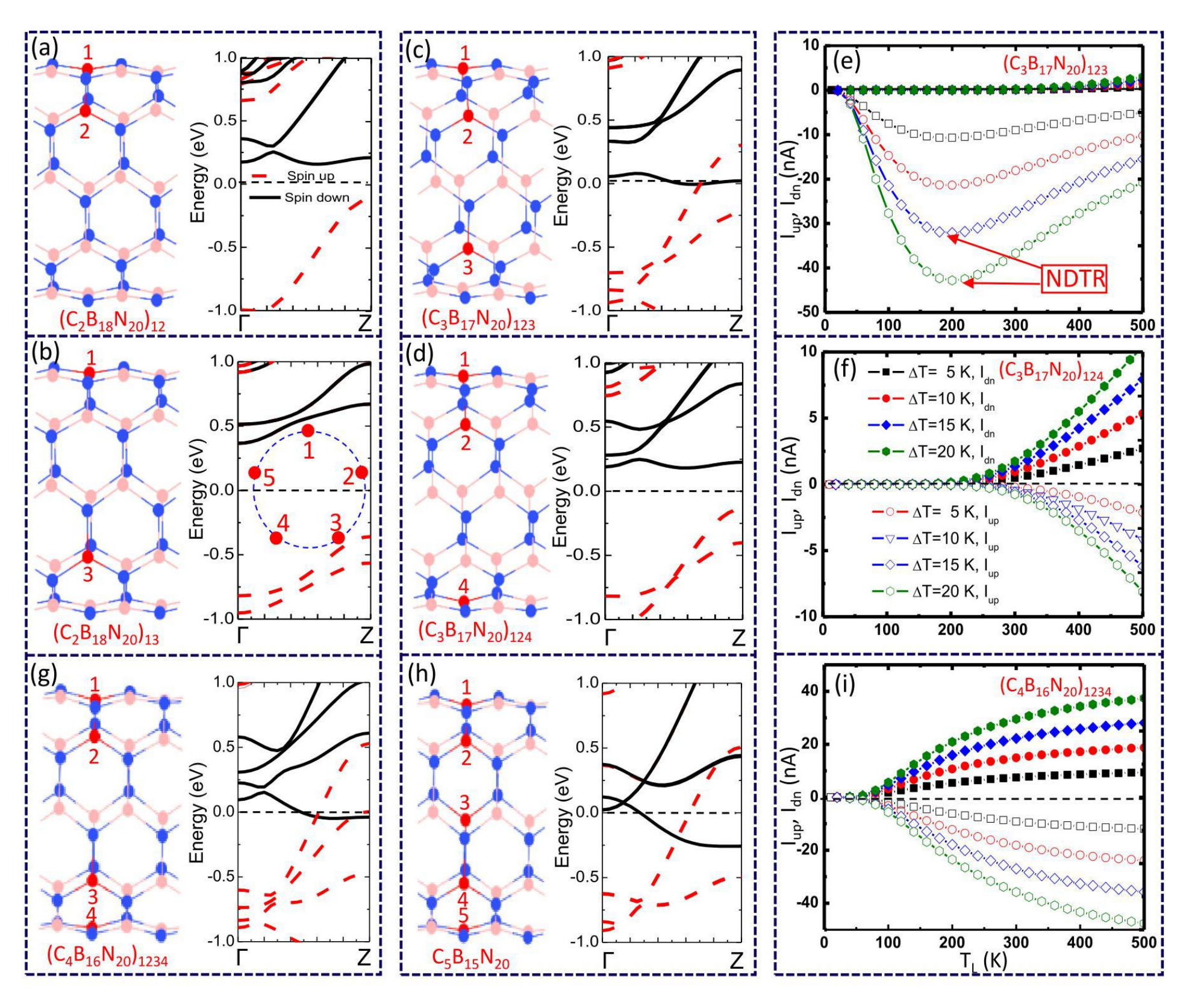}
\caption{\label{fig:wide}  (a)-(b) The schematic structures and the band structures of the two-carbon-substituted armchair (C$_{2}$B$_{18}$N$_{20}$)$_{12}$- and (C$_{2}$B$_{18}$N$_{20}$)$_{13}$-BNNTs. In the left panels, one unit cell is drawn (the same in the below), and composed of 2 carbon atoms, 18 boron atoms and 20 nitrogen atoms. (c)-(d) The nanoribbon structures and the band structures of the three-carbon-substituted armchair (C$_3$B$_{17}$N$_{20}$)$_{123}$- and (C$_{3}$B$_{17}$N$_{20}$)$_{123}$-BNNTs, in which the every unit cell is composed of 3 carbon atoms, 17 boron atoms and 20 nitrogen atoms. (e)-(f) The thermal driven spin-up current ($I_{up}$) and spin-down current ($I_{dn}$) in the magnetic metallic (C$_{3}$B$_{17}$N$_{20}$)$_{123}$-BNNT and the magnetic semiconducting (C$_{3}$B$_{17}$N$_{20}$)$_{124}$-BNNT versus $T_L$, where the temperature gradient $\Delta$\emph{T} is set as 5, 10, 15 and 20 K. (g)-(h) The nanoribbon structures and the band structures of the (C$_{4}$B$_{16}$N$_{20}$)$_{1234}$- and (C$_{5}$B$_{15}$N$_{20}$)-BNNTs. (f) $I_{up}$ and $I_{dn}$ in the magnetic metallic (C$_{4}$B$_{16}$N$_{20}$)$_{124}$-BNNT versus $T_L$.}
\end{figure*}

\begin{figure}
\includegraphics[width=3.35in]{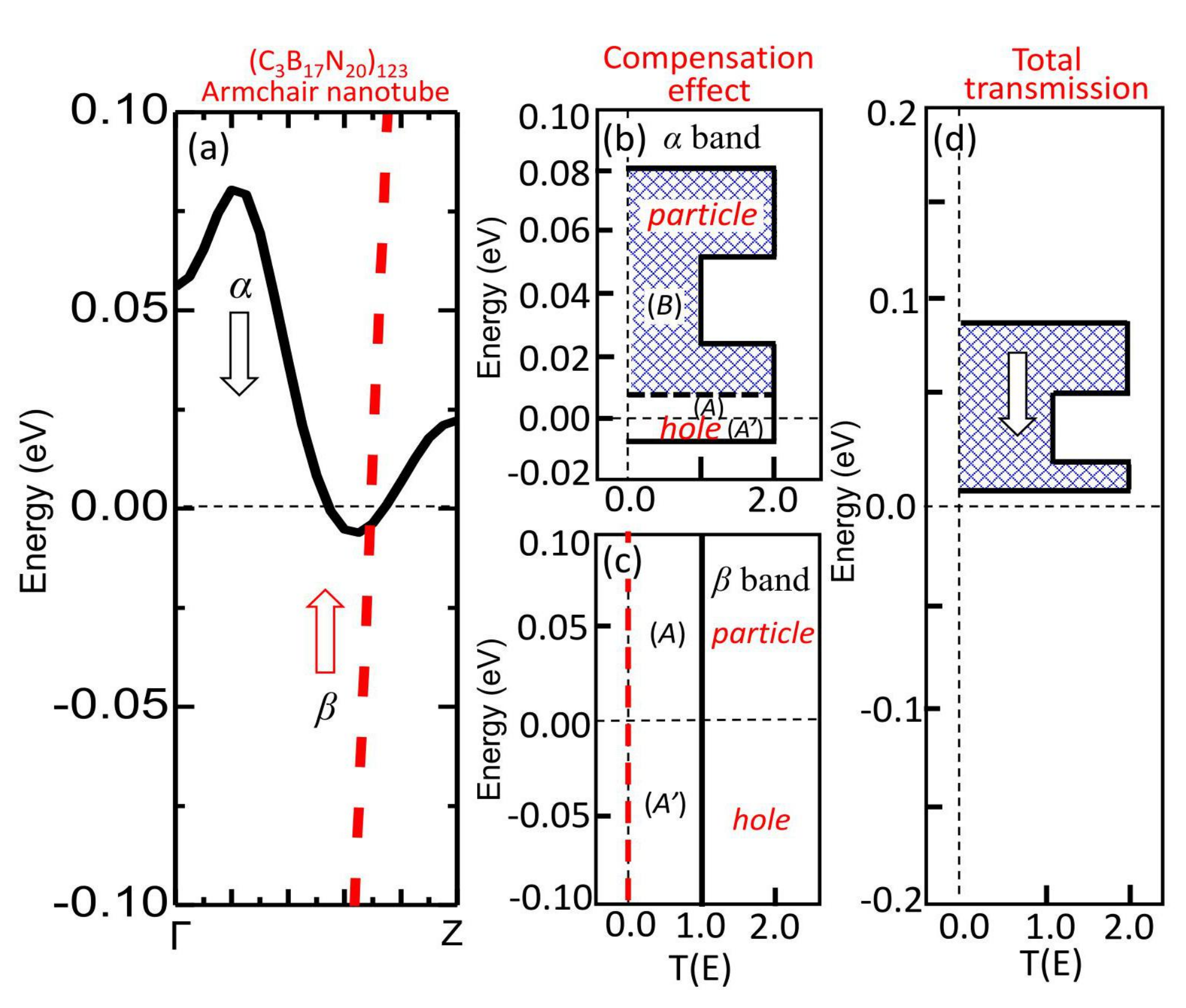}
\caption{\label{fig:wide} (a) The spin-dependent band structures of the armchair (C$_{3}$B$_{17}$N$_{20}$)$_{123}$-BNNT around the Fermi level, where a spin-down band ($\alpha$ band) and a spin-up band ($\beta$ band) are drawn. (b) and (c) The formation of the compensation effect in the electron and hole transmissions of $\alpha$ and $\beta$ bands around the Fermi level. (d) The net transmission of the nanotube in the energy region -0.2 eV $< E - E_F <$ 0.2 eV.}
\end{figure}

The SSE can be further enhanced by the increasing concentration of dopants, i.e., the number of carbon atoms in the BN ring. To test the feasibility, we construct some armchair (C$_n$B$_{20-n}$N$_{20}$)$_{p{\cdot\cdot\cdot}q}$-BNNTs, where the subscripts $p{\cdot\cdot\cdot}q$ refer to doping sites around the BN ring, as plotted in the inset of Fig. 3(b). For $n = 2$, there are only two nonequivalent structures, i.e., (C$_2$B$_{18}$N$_{20}$)$_{12}$- and (C$_2$B$_{18}$N$_{20}$)$_{13}$-BNNTs, as drawn in Figs. 3(a) and 3(b). Both structures are magnetic semiconductors but the latter tends to generate a better SSE, because it possesses more symmetric spin-splitting bands around the Fermi level. For $n = 3$, there are also two different structures, i.e., (C$_3$B$_{17}$N$_{20}$)$_{123}$- and (C$_3$B$_{17}$N$_{20}$)$_{124}$-BNNTs, as drawn in Figs. 3(c) and 3(d). Their bands suggest that the former is a magnetic metal, while the latter is a spin semiconductor, tending to generate a better SSE due to the symmetric spin-up and -down bands near the Fermi level. This is confirmed further by the thermal spin currents plotted in Fig. 3(f), in which we can see that $I_{up}$ and $I_{dn}$ have the opposite signs with the same amplitudes as $T_L > T_{th}$. In comparison with the BNNT with $n$ = 1, both $I_{up}$ and $I_{dn}$ increase largely by three orders of magnitude, due to the deceasing spin-splitting band gap. However, when \emph{n} is increased further (c.f. $n = 4$ and 5), the nanotubes display as magnetic metals (see Figs. 3(g) and 3(h)). It appears that the number of carbon atoms cannot exceed 3, if one wants the armchair BNNTs to have the semiconducting feature with nearly symmetric spin-dependent transmissions around the Fermi level. To grasp further a general rule of the SSE in the magnetic BNNTs with the increasing concentration of carbon dopants, we also study the multiple-carbon-substituted zigzag BNNTs in Fig. S2\cite{46}. The additional numerical results show that the lower the rotational symmetry of the carbon atoms is, the weaker the SSE becomes. Therefore, we get another practical rule that a high rotational symmetry of the doping carbons about the BN ring contributes to generate a better SSE nearly without any charge currents in the BNNTs.

A question arises naturally is that can the metallic and magnetic BNNTs mentioned above generate a good SSE with finite threshold temperatures? To answer this question, we calculate the thermally driven $I_{up}$ and $I_{dn}$ through the metallic (C$_3$B$_{17}$N$_{20}$)$_{123}$- and (C$_4$B$_{16}$N$_{20}$)$_{1234}$-BNNTs plotted in Figs. 3(e) and 3(i), respectively. It is interesting that the former generates a nearly perfect SFE, since $I_{dn}$ increases to finite values with a threshold temperature while $I_{up}$ keeps zero in the entire temperature range. Moreover, $I_{dn}$ versus $T_L$ behaves as a negative differential thermoelectric resistance (NDTR), indicating that it can be designed as particular thermal spin devices with multiple properties. The latter, however, tends to generate a good SSE, since $I_{up}$ and $I_{dn}$ are nearly symmetric about the zero-current axis. More importantly, the SSE occurring in the magnetic and metallic nanotubes has the typical characteristics of spin semiconductors as shown in Fig. 3(f), including the finite threshold temperatures originating from the band gaps.

To understand the SSE and the SFE with finite threshold temperatures occurring in the metallic nanotubes, we turn to analysis their electronic structures and spin-dependent transmissions. Since the carriers around the Fermi level dominate the transport characteristics of a device, we re-plot the band structures of the (C$_3$B$_{17}$N$_{20}$)$_{123}$-BNNT in the energy region -0.1 eV $< E - E_F <$ 0.1 eV in Fig. 4(a), which includes a spin-down band ($\alpha$ band) and a spin-up band ($\beta$ band). The transmission of $\alpha$ band is located in the energy region -0.006 eV $<$ $E - E_F$ $<$ 0.081 eV, and composed of the particle and hole parts. From Eq. (1), the spin currents are determined by the net transmissions. For the example of $\alpha$ band, the hole part $A'$ is cancelled by the electron part $A$, remaining the net transmission region $B$, as shown in Fig. 4(b). For convenience, we define this behavior as compensation effect. Just due to this effect, the net transmission for $\beta$ band is zero around the Fermi level. As a result, in the energy region -0.2 eV $< E - E_F <$ 0.2 eV, the total transmission of the (C$_{3}$B$_{17}$N$_{20}$)$_{123}$-BNNT distributes only in the energy region 0.006 eV $< E - E_F <$ 0.081 eV with spin-down electrons as plotted in Fig. 4(d). As the temperature bias is increased over a threshold temperature, the spin-down electrons are permitted to transport through the device, resulting in the SFE. Moreover, as $T_L$ is increased, the Fermi-Dirac function $f_{L(R)}$ pushes the carriers towards the high-temperature region, which reduces the carriers transporting through the above channel, resulting in the NDTR. Furthermore, the compensation effect can be used to understand the occurrence of the SSE with finite threshold temperatures in the metallic (C$_{4}$B$_{16}$N$_{20}$)$_{1234}$-BNNT, and to explain the generations of the SSE or the SFE in the metallic zigzag BNNTs in the supplemental Fig. S3\cite{46}. It can be concluded that for the magnetic and metallic BNNTs, as the substituted carbon atoms have a high rotational symmetry in the BN ring, the nanotube can generate a good SSE, otherwise tends to a thermal SFE; and the compensation effect makes the SSE or the SFE to hold finite threshold temperatures. This could be regarded as the third practical rule of the magnetic nanotubes to generate the SSE.

\begin{figure}
\includegraphics[width=3.2in]{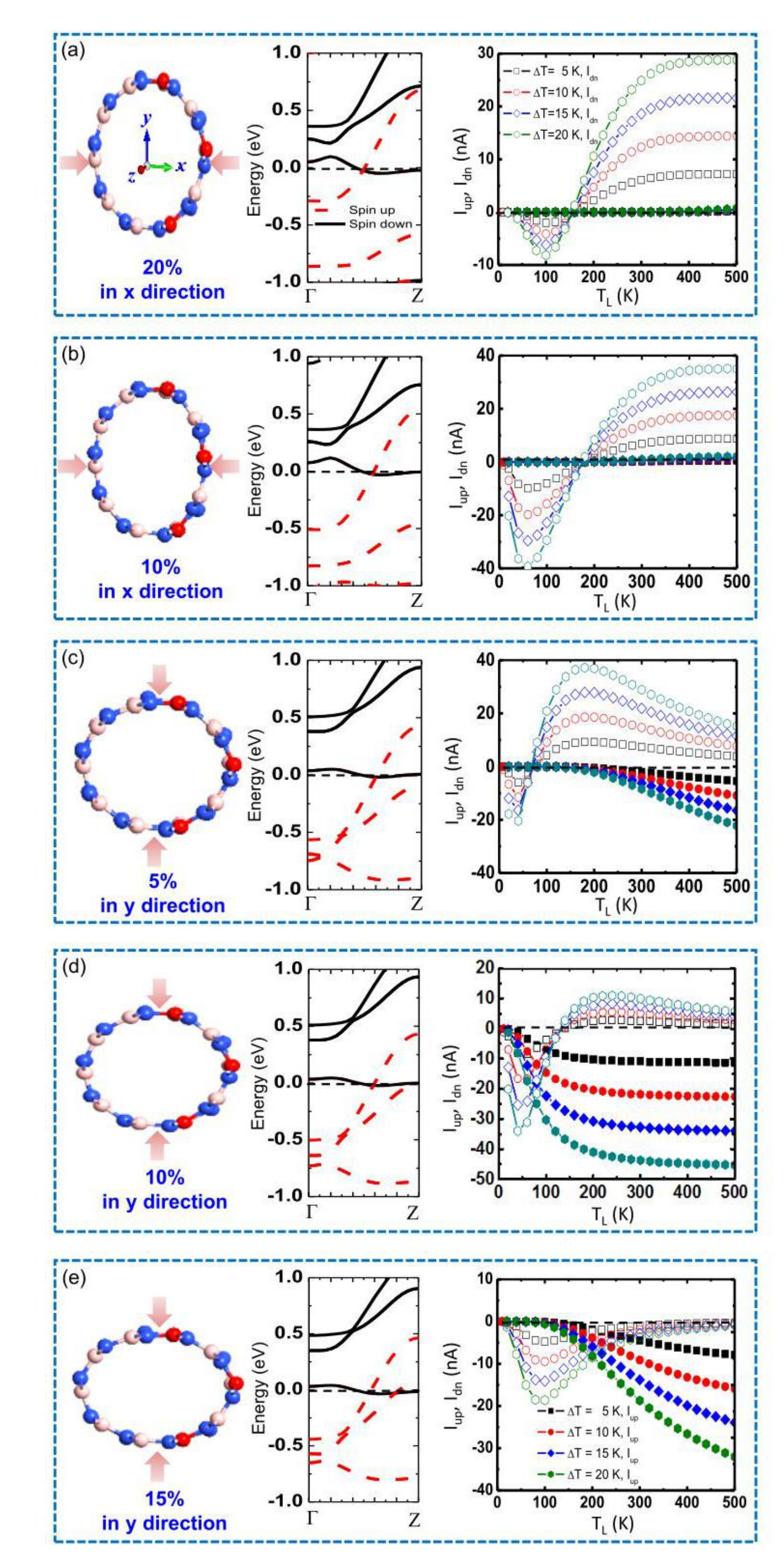}
\caption{\label{fig:wide} The compression strain engineering is used to improve the SSE and the thermal SPE, and to realize the transition between these two effects. The example is adopted as the armchair (C$_3$B$_{17}$N$_{20}$)$_{123}$-BNNT, and in the model, the \emph{x} direction crosses over the carbon with the index $p = 2$ and the \emph{y} direction is along its vertical radical as shown in the inset of figure (a). In the figures (a) and (b), the compression strain is applied in the \emph{x} direction with the proportion values of 20$\%$ and 10$\%$; in the figures (c)-(e), the compression strain is applied in the \emph{y} direction with the proportion values of 5$\%$, 10$\%$ and 15$\%$. The corresponding band structures of these models are plotted in the central panels, and the thermally driven spin-up and spin-down currents ($I_{up}$ and $I_{dn}$) versus $T_L$ are shown in the right panels, where the temperature gradients are adopted in the figures.}
\end{figure}

Towards the practical thermo-spin device applications based on the magnetic BNNTs, an urgent need is to achieve effective mechanisms to improve and control the SSE and the SFE. It is well known that strain engineering has been developed as a well-established approach to enhance the performance of electronic devices \cite{48}, by tuning band structure and carrier mobility of semiconductors \cite{49,50}. Since general nanotubes have a large Yong¡¯s modulus in the axial direction \cite{51,52}, while they are usually soft in the radial direction \cite{53,54}, we try to apply the compression strain in the two orthogonal radial directions of the BNNTs, and take the armchair (C$_{3}$B$_{17}$N$_{20}$)$_{123}$-BNNT as an example. In the model, the \emph{x} direction crosses over the carbon with the index $p = 2$ and the \emph{y} direction is along its vertical radical shown in Fig. 5(a). As the compression strain is applied in the \emph{x} direction, the nearly perfect thermal SFE maintains well, while the spin-down currents are turned to their opposite flow directions, accompanying with the suppressed NDTR peaks as the compression is increased. As the compression strain is applied in the \emph{y} direction, however, the situations are much different. For a low proportion value of 5, apart from the changing flow directions in the spin-down currents, the NDTR effect appears in both transporting directions. Meanwhile, as $T_L$ increases, the spin-up currents increase from zero to form symmetric trends with the spin-down currents. As a result, a SSE nearly without any thermal charge current occurs again, as shown in Fig. 5(c). As the compression strain is increased further, both spin-up and spin-down currents tend to flow in the same directions, behaving as a temperature-dependent SFE (see Fig. 5(e)). These interesting spin transport properties can be understood from the tuning bands (plotted in the central panel) together with the compensation effect. Thus, we believe that the compression strain engineering is an effective route to improve the SSE and the SFE, and to realize other inspiring thermo-spin transports in the magnetic nanotubes. This is another advantage of the magnetic nanotubes to be applied as the thermal spin devices in comparison with nanowire and nanoribbon systems.

Since the spin currents $I_{up}$ and $I_{dn}$ are driven by the temperature gradient in the devices, the magnetic BNNTs would work at finite temperatures. Thus to ensure the occurrence of the SSE and the thermal SFE, the Curie temperature ($T_c$) of the magnetic BNNTs should be larger than $T_{th}$, which is as another urgent need to push the carbon-substituted BNNTs studied here towards the practical device applications. It is noted that $T_{th}$ of the magnetic BNNTs is determined by the spin-splitting band gap. For example, in the one-carbon-substituted (5,5) armchair BNNT drawn in Fig. 1(b), $T_{th}$ is high to 350 K due to the fact that the spin-splitting band gap is large nearly to 0.49 eV. From our supplementary calculations, we find that as another substituted carbon atom is added to the above magnetic BNNT along the transport direction (see the nanotube model drawn in Fig. S1(c)\cite{46}), its spin-splitting band gap narrows down to 0.125 eV and $T_{th}$ decreases nearly to 150 K, while the SSE maintains well as illustrated in Fig. S1(c)\cite{46}. Furthermore, since the magnetic coupling between two nearest carbon atoms in the latter nanotube is enhanced, $T_c$ is increased remarkably. This indicates that to enhance the concentration of the carbon dopants is an effective route to optimize the magnitudes of $T_c$ and $T_{th}$ to ensure the realization of the SSE and the thermal SFE in the magnetic BNNTs.

\section{CONCLUSION}

In summary, we have proposed a new material platform, i.e., the boron-nitrogen nanotubes with carbon substitutions, to realize the robust SSE and the thermal SFE. Through adjusting the rotational symmetry of the doping carbons in the BN ring and enhancing the concentration of carbon dopants in the BNNTs, we have uncovered some practical rules of thermal spin transport phenomena in nanotube systems. As only one boron atom in the unit cell is substituted by the carbon in a BN ring, both the armchair and zigzag BNNTs display as spin semiconductors to generate a robust SSE nearly without any thermal charge current. Moreover, as the concentration of carbon dopants in the unit cell is increased, a high rotational symmetry of carbon atoms contributes to generate a good SSE, otherwise tends to a thermal SFE. Moreover, some metallic BNNTs can also generate the SSE or thermal SFE with finite threshold temperatures, due to the compensation effect of particles and holes around the Fermi level. Furthermore, we find that the compression strain engineering is an effective mechanism to improve and control the SSE and the SFE, to realize the transition between them, and even to obtain other inspiring thermo-spin transport phenomena. Finally, we propose a feasible way to optimize $T_c$ and $T_{th}$ to ensure the generation of the SSE and the SFE in the magnetic BNNTs. These theoretical results will enrich the spin caloritronics and expand the materials candidates to realize robust SSE and other thermal spin effects.

\begin{acknowledgments}
This work is supported by the National Natural Science Foundation of China with grant No. 11774104 and 11274128. Work at UCI was supported by DOE-BES (Grant No. DE-FG02-05ER46237). Computer simulations were partially performed at the U.S. Department of Energy Supercomputer Facility (NERSC).
\end{acknowledgments}


\end{document}